\documentclass[conference]{IEEEtran}
\IEEEoverridecommandlockouts
\usepackage{cite}
\usepackage{algorithmic}
\usepackage{graphicx}
\usepackage{textcomp}
\usepackage{amsmath,amssymb,amsfonts}
\usepackage{flushend}
\usepackage{lipsum}
\usepackage{xcolor}
\usepackage{hyperref} 

\def\BibTeX{{\rm B\kern-.05em{\sc i\kern-.025em b}\kern-.08em
    T\kern-.1667em\lower.7ex\hbox{E}\kern-.125emX}}
\begin{document}
\title{Comparative Evaluation of Transfer Learning for 
Classification of Brain Tumor Using MRI}

\author{\IEEEauthorblockN{Abu Kaisar Mohammad Masum\IEEEauthorrefmark{1},
Nusrat Badhon\IEEEauthorrefmark{2},
S.M. Saiful Islam Badhon\IEEEauthorrefmark{3}, 
Nushrat Jahan Ria\IEEEauthorrefmark{2},\\
Sheikh Abujar\IEEEauthorrefmark{4},
Muntaser Mansur Syed
\IEEEauthorrefmark{1}, and
Naveed Mahmud\IEEEauthorrefmark{1}
}
\IEEEauthorblockA{\IEEEauthorrefmark{1}Electrical Engineering and Computer Science, Florida Institute of Technology, Melbourne, FL, USA}
\IEEEauthorblockA{\IEEEauthorrefmark{2}Department of Computer Science and Engineering, Daffodil International University, Dhaka, Bangladesh}
\IEEEauthorblockA{\IEEEauthorrefmark{3}Department of Information Science, University of North Texas, Denton, TX, USA}
\IEEEauthorblockA{\IEEEauthorrefmark{4}Department of Computer Science, The University of Alabama at Birmingham, Birmingham, AL, USA}
\texttt{\{amasum2022@my., msyed2011@my., nmahmud\}@fit.edu}\\
\texttt{\{nusrat15-12093, nushratria.cse\}@diu.edu.bd}\\
\texttt{{smsaifulislambadhon}@my.unt.edu}\\
\texttt{{sabujar}@uab.edu}\\
[-2.8ex]
}
\maketitle
\begin{abstract}
Abnormal growth of cells in the brain and its surrounding tissues is known as a brain tumor. There are two types, one is benign (non-cancerous) and another is malignant (cancerous) which may cause death. The radiologists' ability to diagnose malignancies is greatly aided by magnetic resonance imaging (MRI). Brain cancer diagnosis has been considerably expedited by the field of computer-assisted diagnostics, especially in machine learning and deep learning. In our study, we categorize three different kinds of brain tumors using four transfer learning techniques. Our models were tested on a benchmark dataset of $3064$ MRI pictures representing three different forms of brain cancer. Notably, ResNet-50 outperformed other models with a remarkable accuracy of $99.06\%$. We stress the significance of a balanced dataset for improving accuracy without the use of augmentation methods. Additionally, we experimentally demonstrate our method and compare with other classification algorithms on the CE-MRI dataset using evaluations like F1-score, AUC, precision and recall.
\end{abstract}


\section{Introduction}
Diseases can be caused by external sources like infections and internal dysfunctions, and they are frequently identified by distinct signs and symptoms. Cancer is regarded as the most dangerous and life-threatening of these illnesses. There are almost $2$ million new cases of brain tumors per year in Bangladesh, which has a population of $165$ million \cite{b1}\cite{b2}.

The type of brain tumor differs depending on the location, size, and shape of the vital organ \cite{b3}. Based on variables like location, patient age, and general health \cite{b4}, whether a brain tumor is benign or malignant affects therapy and prognosis. Moreover, brain tumor incidence is rising in Bangladesh, which may be related to environmental, lifestyle, or hereditary causes. In order to account for malignancy \cite{b5}, the World Health Organization (WHO) divides brain tumors into two classes. The origin, location, and benign or malignant status of the tumor determine its classification. Astrocytomas, oligodendrogliomas, and ependymomas are frequent gliomas that develop from glial cells \cite{b6}.

Brain tumors are diagnosed using a mix of physical examinations, patient history reviews, and medical imaging techniques like CT scans, MRIs, and biopsies\cite{b7}\cite{b8}. Surgery, radiation therapy, chemotherapy, and other medical procedures are all available as treatments for brain tumors in Bangladesh. Radiologists are favoring medical imaging modalities more and more due to their effectiveness and patient safety. Particularly MRI provides rich soft tissue information through intricate multidirectional imaging, making it a key technique for finding brain malignancies. However, the technique has some drawbacks such as the inability to handle huge datasets, automate classification, or handle non-linear relationships between inputs and outputs \cite{b9}.

People suffer every year as a result of misdiagnosis of the kind of tumor in the early stages or unanticipated tumor discoveries during the initial test \cite{b10}. Machine learning techniques have been used by numerous writers to categorize various tumor forms \cite{b11}\cite{b12}\cite{b13}\cite{b14}. Despite substantial advances in picture segmentation and classification, machine learning still has some drawbacks \cite{b15}\cite{b16}. Especially with diverse datasets, traditional machine learning algorithms fail to learn complicated image characteristics and frequently rely on human-engineered features that might not capture all relevant information. In contrast to traditional methods, Convolutional Neural Networks (CNNs) \cite{b17} which are built for image analysis, excel at seeing complex patterns in medical imaging that can be missed. This allows for a more accurate classification of brain tumors.

This study is an effort to improve brain tumor classification by utilizing four well-known pre-trained models such as ResNet50, VGG16, Inception-V3, and MobileNet-V2, and their evaluation scores. These models were polished and trained using a large dataset of brain tumor images. The goal of the study was to better understand performance indicators and the comparative performance of the four pre-trained models in the particular domain of brain tumor categorization. By enabling better diagnosis and treatment planning for people with brain tumors, this research advances the field.

\section{Related Work}
 Guan et al. proposed \cite{b18} a model to improve the visual quality of images by utilizing contrast optimization and nonlinear methods, segmentation and clustering to identify tumors, EfficientNet for feature extraction, refinements to enhance detection accuracy, and data augmentation to avoid over-fitting. It was tested on a FigShare dataset, and the results showed a $98.04\%$ total classification accuracy. 
 
 Another study proposed \cite{b19} the implementation of a deep learning framework that employs transfer learning techniques for the purpose of classifying brain tumors through the analysis of images from a limited dataset using ResNet50, Xception, and MobileNet-V2 CNN architectures. The initial picture dataset was increased by the authors from $253$ to $1516$ images. The MobileNet-V2 architecture achieved the best results with $98.42\%$. Palash Ghosal et al. \cite{b20} utilized a publicly available tumor database in their investigation. The model attained an aggregate accuracy of $89.93\%$ in the absence of data augmentation. After incorporating data augmentation, the model's overall accuracy improved to $93.83\%$. 

The hybrid CNN method is faster than other deep-learning methods while maintaining high classification accuracy. In the same field, Hao Dong et al. proposed \cite{b21} a 2D fully convoluted segmentation network and Sadia et al. \cite{b22} introduced an automated brain tumor detection process on $200$ T2 weighted MRI images. The proposed method surpassed earlier research methodologies, with $98.57\%$ accuracy for Discrete Wavelet Transform (DWT), Principle Component Analysis (PCA), and Kernel Support Vector Machine (KSVM) (IPOL). Support Vector Model(SVM) models were tested with two types of kernels (LIN, IPOL). Using the Stochastic Gradient Descent (SGD) method with a rate of learning of $0.001$ and momentum of $0.9$, three ConvNet models were tested on the TCGA GBM and LGG dataset in another work \cite{b23}. VolumeNet outperformed the competition, achieving maximal accuracy in only $20$ iterations. While both PatchNet and SliceNet performed well in validation, SliceNet was more accurate in training. On the validation set, both the VGGNet and ResNet pre-trained models performed at $85\%$ accuracy.

\section{Methodology}
\begin{figure*}[h!]
 \center
  \includegraphics[width=14cm,height=5cm]{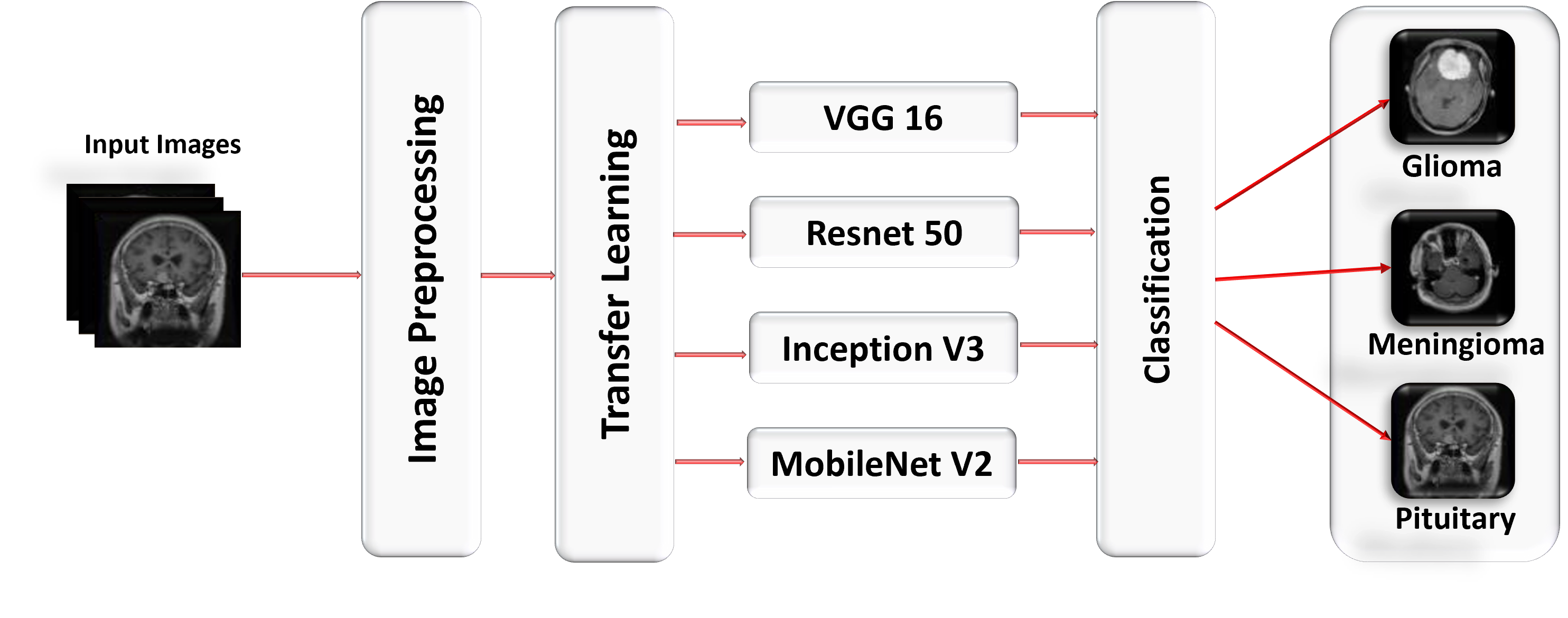}
  \caption{Overview of the MRI images classification using various transfer learning methods}
  \label{Figure 1}
\end{figure*}

The effectiveness of CNN models that have already been trained was thoroughly assessed in this study. ResNet50, Inception-V3, VGG16, and MobileNet-V2 were among the four pre-trained models used. Figure \ref{Figure 1} provides specifics of the study's workflow. An established dataset on brain tumors from Figshare was used to verify the suggested methodology. Preprocessing operations were performed on the dataset, such as resizing, cropping, and scaling, but augmentation techniques were not used. The preprocessed images were then fed into pre-trained CNN models, which extracted pertinent features, which were then supplied into a softmax classifier. According to the study's conclusions, the suggested method may be able to offer a dependable and efficient methodology for classifying brain tumors.

\subsection{Dataset}
The Figshare dataset \cite{b24} contains MRI pictures of brains with different forms of malignancies. Through the Brain Tumor Segmentation (BraTS) competition, a community-driven effort aimed at creating and analyzing cutting-edge algorithms for brain tumor segmentation, these photos were gathered and made available. The collection consists of the MRIs of $285$ patients who are all battling brain tumors, such as gliomas, meningiomas, and pituitary adenomas. The collection includes several MRI scans \cite{b25} for each participant, with each pixel measuring $0.49$ by $0.49$ units. The collection also includes photos of the brain's white matter, gray matter, and cerebrospinal fluid, as well as manually segmented tumor sections \cite{b26}. It contains a total of $1426$ glioma pictures (representing $89$ patients) and $930$ pituitary tumor images (representing $62$ patients), all collected from brain MRI scans.
\begin{figure}[h!]
 \center
  \includegraphics[width=6cm,height=4cm]{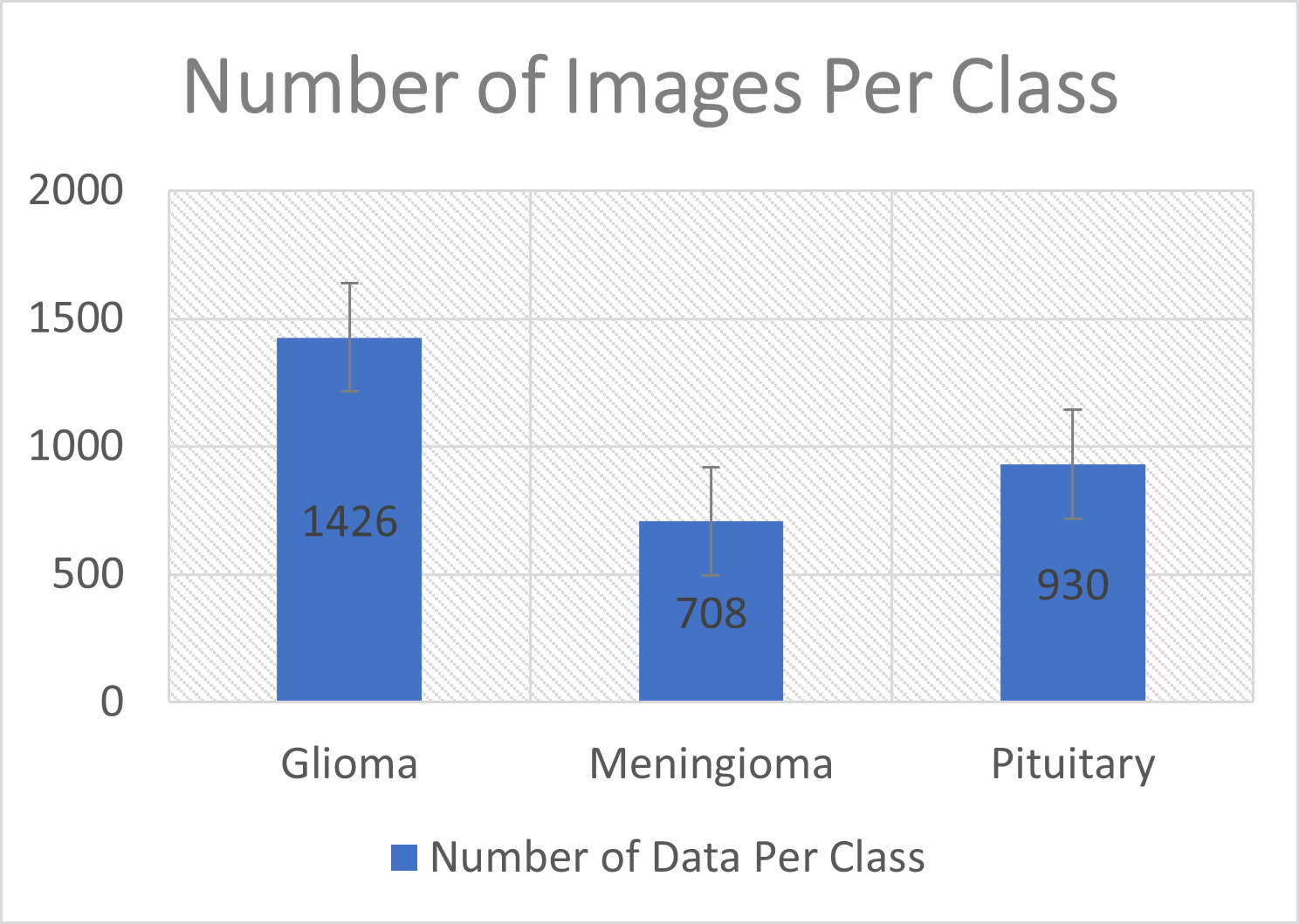}
  \caption{Data Distribution for Each Class.}
  \label{Figure 2}
\end{figure}
\subsection{Dataset Split}
Data used in the experiments were split using two different ratios to acquire the most exact findings. First, a split of $70-15-15$ was used, with training receiving $70\%$ of the budget, validation receiving $15\%$, and testing receiving $15\%$. Following the $80/20$ rule \cite{b27}, the dataset was then divided into $20\%$ for testing and validation and $80\%$ for training. To determine the best split that produced the highest model accuracy, different ratios were tested. Notably, the VGG16 model successfully implemented an $80-20$ split with an exceptional success rate of $99.69\%$ without the need for augmentation methods or difficult pre-processing procedures. Figure \ref{Figure 2} shows the distribution of data used in this study.

\subsection{Image Preprocessing}
 The main goal of this stage was to classify the photos and pre-process them in a way that would make further analysis easier. The photos were loaded using the 'h5py' module, which also extracted crucial information from each file, including tumor border coordinates, image arrays, and labels. By standardizing pixel values between $0$ and $255$, which correspond to the minimum and maximum intensities, a normalization technique was used to assure dataset homogeneity and meaningful comparisons \cite{b28}. A rotation of $-90$ degrees was used to maintain consistency in image orientation. Meningioma, glioma, and pituitary tumors were the three main tumor classes in this study.

\subsection{Dataset Balance}
In collected data, $708$ cases of meningioma, $1426$ cases of glioma, and $930$ cases of pituitary tumors were found. The weights for each class were changed after applying the class weight technique: "Pituitary" was given a weight of $0.716$, "Meningioma" a weight of $1.443$, and "Glioma" a weight of $1.098$. By giving minority classes (meningioma and glioma) higher weights and the majority class (pituitary) a lower weight, this method is intended to reduce dataset imbalance. In order to reduce bias toward the dominant class and improve the model's capacity to accurately classify all tumor types, the objective was to establish a more balanced class representation and prioritize underrepresented classes during training.

\subsection{Transfer Learning}
The application of transfer learning, a machine learning technique, involves adopting a pre-trained model that has been trained for a specific task and using it as the foundation for a related but unfamiliar task \cite{b29}. This approach leverages the knowledge and information gained from the prior task, avoiding the need to start from scratch with a new dataset. Fine-tuning a trained CNN model involves two essential steps such as utilizing the general features learned by the model's initial layers and adapting them for the specific task at hand, along with optimizing the weights of the last few layers to perform well on the new dataset \cite{b30}.

Reduced training time, generally enhanced neural network performance, and the capacity to use fewer data points are the main advantages of transfer learning \cite{b31}. The decoding of picture features relies heavily on convolutional layer architectures. During the convolution process, CNNs use a variety of detectors, including edge and corner detectors, also known as filters. These filters are applied to the edges and vertices of the image, producing convolved copies of the original image that are then used to create feature maps. Through a sequence of processes, convolution seeks to extract information from an image, including corners and edges. Following convolution layers process the high-level image features that these map features capture \cite{b32}. 

\paragraph{VGG16}
VGG16 has grown in popularity among CNN architectures and is frequently used as a fundamental framework for more complex CNN models built via transfer learning. It is the best option for creating complex CNNs because of its deep yet simple structure \cite{b33}. The design, with a total of $16$ layers and $224224$ GB of input images, consists mostly of convolutional layers, followed by fully linked layers \cite{b34}. Three fully connected layers, $13$ convolutional layers, five max-pooling layers, and two dropout layers make up the architecture. $64$ filters with a $3$x$3$ size and a stride of 1 are present in each of the first two convolutional layers. The number of filters in subsequent convolution layers rises to $512$ before being followed by max-pooling layers with a $2$x$2$ kernel and a stride of $2$. Three completely connected layers that each use ReLU activation and has $4096$, $4096$, and $1000$ units each are used to channel the output of the final max pooling layer.

\paragraph{ResNet50}
ResNet is a subset of the larger CNN family, and it has a typical architecture with 48 convolutional layers, one MaxPool layer, and one average pooling layer that is frequently employed in a variety of computer vision applications. ResNet-34 has $34$ weighted layers in its initial form. $50$ layers make up the ResNet-50 design, which also includes a max pooling layer with a 2-stride and a $7x7$ kernel convolution. It also has $512$ cores, $11$x$11$ cores, and layers with kernel sizes of $9$x$9$, $12$x$12$, $18$x$18$, and $9$x$9$ \cite{b35}. The softmax activation function is used in the last layer, which is fully linked and averages a pool of $1000$ neurons.

\paragraph{Inception-V3}
A sophisticated convolutional neural network with 48 layers makes up the Inception-V3 model. The vast ImageNet collection, which contains more than one million unique photographic images, served as its pre-training set. By achieving $78.1\%$ on the ImageNet dataset \cite{b36}, Inception-V3 exhibits astounding accuracy. Inception-V3 introduces improvements to its forebear within the Inception architecture to improve computing efficiency. To lighten the processing load, two $3$x$3$ convolutional layers have been substituted. Deep neural networks' efficiency is increased by asymmetric convolutions like $n$x$1$, and convergence is aided by auxiliary classifiers. Auxiliary classifiers initially had little effect on training but eventually displayed greater accuracy. Expanding the dimensions of the network filter activation by combining convolution and pooling blocks simultaneously reduces grid size.

\paragraph{MobileNet-V2}

The bottleneck layer residual block in MobileNet-V2 is made up of limited layers, and the structure is recursive. In order to filter features in the intermediate expansion layer, simple depth-wise convolutions are used. Input-filtering depthwise convolution, $1$x$1$ pointwise convolution, and another depthwise convolution are the three convolutional layers that make up the basic bottleneck residual block in MobileNet-V2, which still uses depthwise separable convolutions \cite{b37}. The default expansion factor for the expansion layer is $6$, and there are more output channels than input channels. 

In contrast to the input and output tensors, which are low-dimensional, this block filters using a high-dimensional tensor. In order to disperse gradients throughout the network, MobileNet-V2 includes a crucial residual connection. It uses batch normalization in each layer, ReLU6 activation across all layers, and does not use activation in the projection layer. A conventional $1$x$1$ convolution, a global average pooling layer, and a classification layer round out the network's last $17$ building components. Instead of an extension layer, the first block uses a normal $3$x$3$ convolution with $32$ channels, setting it apart from the remainder of the network's structure. After laying a more straightforward basis, the network topology of MobileNet V2 enables the construction of a more effective network.

\section{Experimental Analysis}

In this experiment we used a dataset of $3064$ samples and modified the transfer learning models by changing the number of neurons in the fully-connected layer to correspond to the three classes in the dataset. Classification models are evaluated using criteria including accuracy, precision, recall, and F1-score. Each network is trained over $10$ epochs with a batch size of $16$, using the Adam optimization method and the Sparse Categorical Cross entropy loss function. Studies using various ratios show that the $80:20$ ratio, which uses $80\%$ of the data for training and $20\%$ for testing and validation, produces the best reliable results. The study uses Python 3 and packages including TensorFlow, Keras, and scikit-learn for effective deep learning training within the Google Colab environment, along with a Core i5 processor, $16$ GB of RAM, and a T4 GPU. Meningioma ($708$ cases), glioma ($1426$ instances), and pituitary ($930$ instances) made up the majority of the brain tumor categories after the dataset had been balanced.

\subsection{Result Before Balancing Dataset}
Prior to performing dataset balancing, the distribution of brain tumor types was as follows: meningioma ($708$ instances), glioma ($1426$ instances), and pituitary ($930$ instances). 
\paragraph{ResNet50}
The ResNet50 model demonstrated an accuracy rate of $94\%$ and exhibited favorable precision, recall, and F1-score metrics for the glioma, meningioma, and pituitary classes, indicating promising performance in classifying these types of brain tumors. Nevertheless, the system encountered difficulties in accurately detecting meningioma, exhibiting a lower recall rate of $83\%$. The Pituitary class demonstrated high performance, achieving a precision score of $1.00$ and a recall score of $0.99$.

\paragraph{VGG16}
The VGG16 model demonstrated a comparable accuracy rate of $95\%$ and exhibited consistent precision, recall, and F1-score metrics across all classes. The results demonstrated favorable performance in the case of glioma and pituitary tumors, although with a slightly diminished recall rate of $0.83$ for meningioma.

\paragraph{Inception-V3}
The Inception-V3 model demonstrated a classification accuracy of $92\%$and displayed a comparatively lower recall rate of $61\%$ specifically for meningioma. Nevertheless, it demonstrated great precision and recall rates for pituitary tumors, thereby showcasing exceptional proficiency in the identification of this specific tumor category.

\paragraph{MobileNet-V2}
The MobileNet-V2 model attained a $96\%$ accuracy rate and demonstrated consistent precision, recall, and F1-score measurements across all categories. The performance of the model was satisfactory for glioma and pituitary tumors, although it exhibited a slightly lower recall rate for meningioma cases.
When comparing the ResNet50 and VGG16 models, it was observed that both models exhibited comparable accuracies and demonstrated effective performance in distinguishing between glioma and pituitary tumors. However, the ResNet50 model encountered difficulties in accurately classifying meningiomas. The results of the study indicate that Inception-V3 exhibited a lower recall rate in detecting meningioma, whereas MobileNet-V2 consistently demonstrated high performance across all categories.

\subsection{Result After Balancing Dataset}
The performance of the ResNet50, VGG16, Inception-V3, and MobileNet-V2 models in classifying brain tumor types exhibited significant improvement following the balancing of the dataset.
\paragraph{ResNet50} 
After applying balancing techniques, the ResNet50 model demonstrated a remarkable accuracy of $99\%$. Notably, it exhibited impeccable precision and recall for the glioma and meningioma classes. Furthermore, the pituitary class achieved a precision of $0.98$ and a recall of $1.00$. This signifies a significant improvement in performance across all classes when compared to the previous outcomes.
\paragraph{VGG16}
After performing dataset balancing, VGG16 demonstrated a notable accuracy rate of $97\%$. Furthermore, it exhibited consistently high precision and recall values for the glioma and pituitary classes. The model exhibited a precision value of $0.93$ and a recall value of $0.93$ specifically for the classification of meningioma. Despite the absence of notable alterations in the performance metrics when compared to the previous outcomes, the utilization of a balanced dataset contributed to enhanced reliability and precision in the predictions.

\paragraph{Inception-V3}
After applying balancing techniques, Inception-V3 demonstrated a notable accuracy rate of $96\%$. Furthermore, the precision and recall metrics for all classes exhibited significant enhancements when compared to the previous findings. The precision and recall metrics for the glioma class were found to be $0.98$ and $0.94$, respectively. Similarly, the meningioma class exhibited precision and recall values of $0.91$ and $0.95$, respectively. Notably, the pituitary class demonstrated impeccable precision and recall.

\paragraph{MobileNet-V2}
After the process of dataset balancing, MobileNet-V2 demonstrated exceptional performance, attaining a remarkable accuracy rate of $99\%$. The results of the model exhibited excellent precision and recall across all classes. Specifically, the glioma class achieved a precision of $1.00$ and a recall of $0.98$, the meningioma class demonstrated a precision of $0.96$ and a recall of $1.00$, while the pituitary class maintained perfect precision and recall. 

\subsection{Comparative Study with State-of-the-Art}

We have balanced the dataset that obtained good results for precision, recall, and F1-score in different classes of VGG16, ResNet50, MobileNet-V2, and Inception-V3 algorithms. The Pituitary class has maximum values of Precision, Recall, and F1-score for these algorithms. As well as the glioma and Meningioma classes also showed adequate results for the classification evaluation. The accuracy for each transfer learning algorithm suggested that ResNet50 provides ($99.06\%$) the highest result for our balanced dataset. In Tables \ref{tab:1}, \ref{tab:2}, \ref{tab:3}, and \ref{tab:4} we provide specific information about each model's performance.

\begin{table}[!h]
    \centering
    \caption{Class- Specific Evaluation for VGG16 \tiny(after dataset balance)}
    \begin{tabular}{|p{1.5cm}||p{1.2cm}|p{1.2cm}|p{1.2cm}|}
    \hline
    VGG16 & Precision & Recall & F1-Score\\
     \hline
     \hline
      Glioma &0.98&0.97&0.97\\
      Meningioma &0.93&0.93&0.93 \\
      Pituitary&0.98&1.00&0.99\\
     \hline
     \hline
     Accuracy&&$97.19\%$&\\
     \hline
    \end{tabular}
    \label{tab:1}
\end{table}
\begin{table}[!h]
    \centering
    \caption{Class- Specific Evaluation for ResNet50 \tiny(after dataset balance)}
    \begin{tabular}{|p{1.5cm}||p{1.2cm}|p{1.2cm}|p{1.2cm}|}
    \hline
    ResNet50 & Precision & Recall & F1-Score\\
     \hline
     \hline
      Glioma &1.00&1.00&1.00\\
      Meningioma &1.00&0.98&0.99 \\
      Pituitary&0.98&1.00&0.99\\
     \hline
     \hline
     Accuracy&&$99.06\%$&\\
     \hline
    \end{tabular}
    \label{tab:2}
\end{table}
\begin{table}[!h]
    \centering
    \caption{Class-Specific Evaluation for Inception-V3 \tiny(after dataset balance)}
    \begin{tabular}{|p{1.5cm}||p{1.2cm}|p{1.2cm}|p{1.2cm}|}
    \hline
    Inception-V3 & Precision & Recall & F1-Score\\
     \hline
     \hline
      Glioma &0.98&0.94&0.96\\
      Meningioma &0.91&0.95&0.93 \\
      Pituitary&0.98&1.00&0.99\\
     \hline
     \hline
     Accuracy&&$95.31\%$&\\
     \hline
    \end{tabular}
    \label{tab:3}
\end{table}
\begin{table}[!h]
    \centering
    \caption{Class-Specific Evaluation for MobileNet-V2  \tiny(after dataset balance)}
    \begin{tabular}{|p{1.7cm}||p{1.2cm}|p{1.2cm}|p{1.2cm}|}
    \hline
    MobileNet-V2 & Precision & Recall & F1-Score\\
     \hline
     \hline
      Glioma &1.00&0.98&0.99\\
      Meningioma &0.96&1.00&0.98 \\
      Pituitary&1.00&1.00&1.00\\
     \hline
     \hline
     Accuracy&&$98.75\%$&\\
     \hline
    \end{tabular}
    \label{tab:4}
\end{table}
\
To represent a thorough comparison between the proposed method and the currently available state-of-the-art methods we give a short summary of our findings in Table \ref{tab:5}. The findings were also compared with previous study. By employing the suggested model, it was observed that state-of-the-art outcomes were achieved in our work.  
\begin{table}[!h]
    \centering
    \caption{Comparison with the previous State-of-the-Art works}
    \begin{tabular}{|p{1cm}||p{0.8cm}|p{1.1cm}|p{1.6cm}|p{0.8cm}|p{0.8cm}|}
    \hline
    Paper&Image Size&Augmen tation&Proposed Method&Training&Max Acc\\
     \hline
     \hline
      [4] &3064 \newline and \newline 253 &Yes&GoogleNet \newline GoogleNet \newline with \newline SVM &$-$&$93.1\%$\newline$98.1\%$ \\
     \hline
      [9] &3064&Yes&ResNet50&$70\%$\newline$98\%$\newline$75\%$&$80\%$\newline$98\%$\newline$99\%$ \\
     \hline
      [20] &3064&Yes&ReseNet101&$-$&$93.83\%$ \\
      \hline
       [30] &2475&No& MobileNet-V2&$80\%$&$94\%$ \\
     \hline
      [39] &3064&No&AlexNet\newline VGG16\newline VGG19&$25\%$\newline$50\%$\newline$75\%$ & $89.95\%$\newline$96.65\%$\newline$94.82\%$ \\
     \hline
     [40] &3064&Yes&VGG19&$50\%$&$90\%$ \\
     \hline
     [41] &3064&Yes&AlexNet\newline GoogleNet\newline VGG16 &$-$&$98.69\%$ \\
     \hline
     [42] &253&$-$&ResNet50\newline Xeption \newline MobileNet-V2 &$80\%$&$77.63\%$\newline$97.35\%$\newline$98.24\%$ \\
     \hline
     [43] &3064&$-$&CNN&$-$&$84.19\%$\\
     \hline
     \hline
     Proposed &2100&No&ResNet50\newline MobileNet-V2 \newline VGG16 \newline Inception-V3 &$80\%$&$99.06\%$\newline$98.75\%$\newline$97.19\%$ \newline$95.31\%$\\
     \hline
     
    \end{tabular}
    \label{tab:5}
\end{table}

\section{Conclusion}
This study's main objective was to use transfer learning techniques for classifying brain tumors. Four pre-trained models ResNet50, VGG16, MobileNet-V2, and Inception-V3 were compared to find which one was the most accurate. The study also looked into the effects of dataset balancing on classification accuracy without using augmentation methods. After dataset balancing, the results showed that ResNet50 performed better than the other models, reaching the maximum accuracy rate of $99.06\%$. Although MobileNet-V2 initially demonstrated the maximum accuracy of $96.88\%$, additional gains were seen with dataset balancing. Future research should examine the possible advantages of augmentation approaches in improving model accuracy and resilience, even if this work did not use them. Additionally, researching different pre-trained models or customized architectures for classifying brain tumors may provide fresh perspectives and enhance classification efficiency in general. To improve accuracy, one may additionally take into account approaches like active learning, Generative Adversarial Networks (GANs), and sophisticated deep learning architectures.

\vspace{12pt}


\begin{thebibliography}{00}
\bibitem{b1}
Suresh H, Jamil S, Padhi BK, Hossain MJ. Thalassemia prevention: Religious and cultural barriers to premarital screening in Bangladesh. Health Science Reports. 2023 Apr;6(4).
\bibitem{b2} 
Hussain SM. Comprehensive update on cancer scenario of Bangladesh. South Asian journal of cancer. 2013 Oct;2(04):279-84.
\bibitem{b3} 
Zhuang F, Qi Z, Duan K, Xi D, Zhu Y, Zhu H, Xiong H, He Q. A comprehensive survey on transfer learning. Proceedings of the IEEE. 2020 Jul 7;109(1):43-76.
\bibitem{b4}
Rasool M, Ismail NA, Boulila W, Ammar A, Samma H, Yafooz WM, Emara AH. A hybrid deep learning model for brain tumour classification. Entropy. 2022 Jun 8;24(6):799.
\bibitem{b5}
\url{https://www.aans.org/en/Patients/Neurosurgical-Conditions-and-Treatments/Brain-Tumors}, Accessed: Sept. 2023
\bibitem{b6}
\url{https://www.cancer.org.au/cancer-information/types-of-cancer/brain-cancer},  Accessed: Sept. 2023
\bibitem{b7}
Sharma K, Kaur A, Gujral S. Brain tumor detection based on machine learning algorithms. International Journal of Computer Applications. 2014 Jan 1;103(1).
\bibitem{b8}
Lavanyadevi R, Machakowsalya M, Nivethitha J, Kumar AN. Brain tumor classification and segmentation in MRI images using PNN. In2017 IEEE international conference on electrical, instrumentation and communication engineering (ICEICE) 2017 Apr 27 (pp. 1-6). IEEE.
\bibitem{b9}
Ismael SA, Mohammed A, Hefny H. An enhanced deep learning approach for brain cancer MRI images classification using residual networks. Artificial intelligence in medicine. 2020 Jan 1;102:101779.
\bibitem{b10}
Vidyarthi A, Mittal N. Performance analysis of Gabor-Wavelet based features in classification of high grade malignant brain tumors. In2015 39th National Systems Conference (NSC) 2015 Dec 14 (pp. 1-6). IEEE.
\bibitem{b11}
Usman, K. and Rajpoot, K., 2017. Brain tumor classification from multi-modality MRI using wavelets and machine learning. Pattern Analysis and Applications, 20, pp.871-881.
\bibitem{b12}
Polly FP, Shil SK, Hossain MA, Ayman A, Jang YM. Detection and classification of HGG and LGG brain tumor using machine learning. In2018 International Conference on Information Networking (ICOIN) 2018 Jan 10 (pp. 813-817). IEEE.
\bibitem{b13}
Sachdeva J, Kumar V, Gupta I, Khandelwal N, Ahuja CK. Multiclass brain tumor classification using GA-SVM. In2011 Developments in E-systems Engineering 2011 Dec 6 (pp. 182-187). IEEE.
\bibitem{b14}
Vaishnavee KB, Amshakala K. An automated MRI brain image segmentation and tumor detection using SOM-clustering and Proximal Support Vector Machine classifier. In2015 IEEE international conference on engineering and technology (ICETECH) 2015 Mar 20 (pp. 1-6). IEEE.
\bibitem{b15}
Swati ZN, Zhao Q, Kabir M, Ali F, Ali Z, Ahmed S, Lu J. Brain tumor classification for MR images using transfer learning and fine-tuning. Computerized Medical Imaging and Graphics. 2019 Jul 1;75:34-46.
\bibitem{b16}
Rehman A, Naz S, Razzak MI, Akram F, Imran M. A deep learning-based framework for automatic brain tumors classification using transfer learning. Circuits, Systems, and Signal Processing. 2020 Feb;39:757-75.
\bibitem{b17}
Yan D, Wu S, Sami MT, Almudaifer A, Jiang Z, Chen H, Rangaprakash D, Deshpande G, Ma Y. Improving Brain Dysfunction Prediction by GAN: A Functional-Connectivity Generator Approach. In2021 IEEE International Conference on Big Data (Big Data) 2021 Dec 15 (pp. 1514-1522). IEEE.
\bibitem{b18}
Guan Y, Aamir M, Rahman Z, Ali A, Abro WA, Dayo ZA, Bhutta MS, Hu Z. A framework for efficient brain tumor classification using MRI images.
\bibitem{b19}
Arbane M, Benlamri R, Brik Y, Djerioui M. Transfer learning for automatic brain tumor classification using MRI images. In2020 2nd International Workshop on Human-Centric Smart Environments for Health and Well-being (IHSH) 2021 Feb 9 (pp. 210-214). IEEE.
\bibitem{b20}
Ghosal P, Nandanwar L, Kanchan S, Bhadra A, Chakraborty J, Nandi D. Brain tumor classification using ResNet-101 based squeeze and excitation deep neural network. In2019 Second International Conference on Advanced Computational and Communication Paradigms (ICACCP) 2019 Feb 25 (pp. 1-6). IEEE.
\bibitem{b21}
Dong H, Yang G, Liu F, Mo Y, Guo Y. Automatic brain tumor detection and segmentation using U-Net based fully convolutional networks. InMedical Image Understanding and Analysis: 21st Annual Conference, MIUA 2017, Edinburgh, UK, July 11–13, 2017, Proceedings 21 2017 (pp. 506-517). Springer International Publishing.
\bibitem{b22}
Alam S, Abdullah M, Khan FN, Ullah AA, Rahi MM, Alam MA. An efficient image processing technique for brain tumor detection from MRI images. In2019 IEEE Asia-Pacific Conference on Computer Science and Data Engineering (CSDE) 2019 Dec 9 (pp. 1-6). IEEE.
\bibitem{b23}
Banerjee S, Mitra S, Masulli F, Rovetta S. Brain tumor detection and classification from multi-sequence MRI: study using ConvNets. InBrainlesion: Glioma, Multiple Sclerosis, Stroke and Traumatic Brain Injuries: 4th International Workshop, BrainLes 2018, Held in Conjunction with MICCAI 2018, Granada, Spain, September 16, 2018, Revised Selected Papers, Part I 4 2019 (pp. 170-179). Springer International Publishing.
\bibitem{b24}
\url{https://figshare.com/articles/dataset/brain_tumor_dataset/1512427},  Accessed: Sept. 2023
\bibitem{b25}
Ghaffari M, Sowmya A, Oliver R. Automated brain tumor Segmentation using multimodal brain scans, a survey based on models submitted to the BraTS 2012-18 challenges. IEEE Rev. Biomed. Eng.:1.
\bibitem{b26}
Ayomide KS, Aris TN, Zolkepli M. Improving Brain Tumor Segmentation in MRI Images through Enhanced Convolutional Neural Networks. International Journal of Advanced Computer Science and Applications. 2023;14(4).
\bibitem{b27}
T. Nisonger, "The “80/20 Rule” and Core Journals," Serials Librarian - SERIALS LIBR, vol. 55, pp. 62-84, 2008. 
\bibitem{b28}
Swati ZN, Zhao Q, Kabir M, Ali F, Ali Z, Ahmed S, Lu J. Brain tumor classification for MR images using transfer learning and fine-tuning. Computerized Medical Imaging and Graphics. 2019 Jul 1;75:34-46.
\bibitem{b29}
Deepak S, Ameer PM. Brain tumor classification using deep CNN features via transfer learning. Computers in biology and medicine. 2019 Aug 1;111:103345.
\bibitem{b30}
Arfan TH, Hayaty M, Hadinegoro A. Classification of brain tumours types based on MRI images using mobilenet. In 2021 2nd International Conference on Innovative and Creative Information Technology (ICITech) 2021 Sep 23 (pp. 69-73). IEEE.
\bibitem{b31}
Mehrotra R, Ansari MA, Agrawal R, Anand RS. A transfer learning approach for AI-based classification of brain tumors. Machine Learning with Applications. 2020 Dec 15;2:100003.
\bibitem{b32}
Liu YH. Feature extraction and image recognition with convolutional neural networks. InJournal of Physics: Conference Series 2018 Sep 1 (Vol. 1087, p. 062032). IOP Publishing.
\bibitem{b33}
Liu S, Deng W. Very deep convolutional neural network based image classification using small training sample size. In2015 3rd IAPR Asian conference on pattern recognition (ACPR) 2015 Nov 3 (pp. 730-734). IEEE.
\bibitem{b34}
Rehman A, Naz S, Razzak MI, Akram F, Imran M. A deep learning-based framework for automatic brain tumors classification using transfer learning. Circuits, Systems, and Signal Processing. 2020 Feb;39:757-75.
\bibitem{b35}
He K, Zhang X, Ren S, Sun J. Deep residual learning for image recognition. InProceedings of the IEEE conference on computer vision and pattern recognition 2016 (pp. 770-778).
\bibitem{b36}
Szegedy C, Vanhoucke V, Ioffe S, Shlens J, Wojna Z. Rethinking the inception architecture for computer vision. InProceedings of the IEEE conference on computer vision and pattern recognition 2016 (pp. 2818-2826).
\bibitem{b37}
Howard AG, Zhu M, Chen B, Kalenichenko D, Wang W, Weyand T, Andreetto M, Adam H. Mobilenets: Efficient convolutional neural networks for mobile vision applications. arXiv preprint arXiv:1704.04861. 2017 Apr 17.
\bibitem{b38}
Kingma DP, Ba J. Adam: A method for stochastic optimization. arXiv preprint arXiv:1412.6980. 2014 Dec 22.
\bibitem{b39}
Swati ZN, Zhao Q, Kabir M, Ali F, Ali Z, Ahmed S, Lu J. Brain tumor classification for MR images using transfer learning and fine-tuning. Computerized Medical Imaging and Graphics. 2019 Jul 1;75:34-46.
\bibitem{b40}
Sajjad M, Khan S, Muhammad K, Wu W, Ullah A, Baik SW. Multi-grade brain tumor classification using deep CNN with extensive data augmentation. Journal of computational science. 2019 Jan 1;30:174-82.
\bibitem{b41}
Rehman A, Naz S, Razzak MI, Akram F, Imran M. A deep learning-based framework for automatic brain tumors classification using transfer learning. Circuits, Systems, and Signal Processing. 2020 Feb;39:757-75.
\bibitem{b42}
Arbane M, Benlamri R, Brik Y, Djerioui M. Transfer learning for automatic brain tumor classification using MRI images. In2020 2nd International Workshop on Human-Centric Smart Environments for Health and Well-being (IHSH) 2021 Feb 9 (pp. 210-214). IEEE.
\bibitem{b43}
Abiwinanda N, Hanif M, Hesaputra ST, Handayani A, Mengko TR. Brain tumor classification using convolutional neural network. InWorld Congress on Medical Physics and Biomedical Engineering 2018: June 3-8, 2018, Prague, Czech Republic (Vol. 1) 2019 (pp. 183-189). Springer Singapore.
\end{thebibliography}
\end{document}